# The noisy image retrieval for fluorescence digital microscopy.

A.Yu.Okulov ∗ *.

*P.N.Lebedev Physical Institute of Russian Academy of Sciences

Leninsky prospect 53, 119991 Moscow, Russia , e_mail: okulov@sci.lebedev.ru

∗ Department of Physics, University of Coimbra,

Coimbra 3000, Portugal

**Abstract.**

The efficient optical tool for elimination of the phase and amplitude distortions produced by imperfectness of the optical elements in microscope lightpath considered. This robust procedure described by simple theoretical model proved to be successful to repair the noisy images with noise to signal ratio close to 100. It is shown that Van Cittert - Zernike theorem provides adequate description of imperfect microscope operating under both coherent and incoherent illumination conditions and having turbid media in object and intermediate planes.

**Introduction.**

The optical images in fluorescence microscopy are affected by a number of factors: among them are the phase and amplitude distortions in the object plane or in the intermediate planes, produced by inhomogeneous flow or turbid particles inside the liquid, surrounding the object. In the typical experimental setup (fig.1) for both trans- and epi - fluorescence microscopy the light from spatially coherent (laser) or incoherent (gas-discharge lamp, like tungsten or xenon one) source passes through the object and acquires some amplitude and phase information concerning the object itself and the properties of the microscope components.

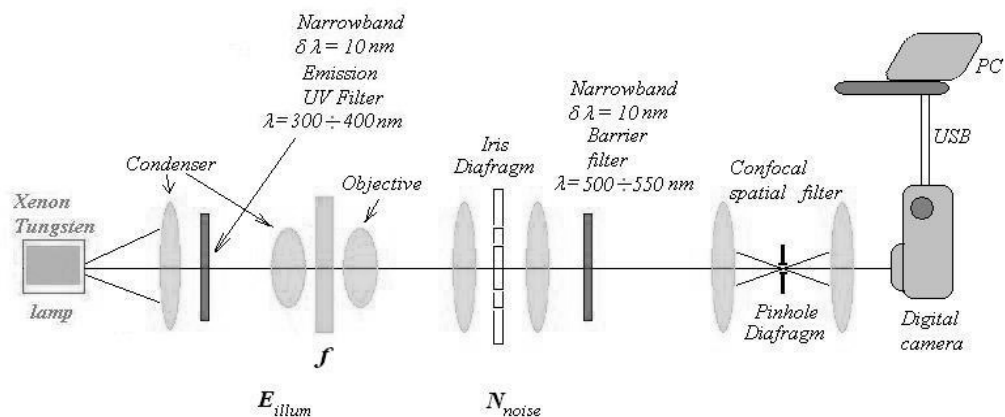

**Fig.1**

These optical components, forming image, namely lenses and mirrors, introduce also both the phase distortions, caused by imperfectness of the surfaces, and amplitude distortions, induced by corrupted reflectivity or dense concentration of a dust particles. Thus in addition to diffractive spread of ideal microscope response function there is a set of additional factors reducing the image quality. These factors are modelled below as random noise field $N(\vec{r})$.

It is well known the quality of image could be substantially enhanced by spatial filtering on the purely "hardware" level, introducing spatial filters/3/, on purely "software" level, by making some digital manipulations with arrays of numbers, produced by detection of the image via analog or digital devices or methods of adaptive optics, which are in fact the combination of two methods, mentioned above, for example by using some deformable optical elements, controlled by computer through specially developed feedback. The goal of the present paper is to describe practically convinient and robust setup for noise elimination from microscopical image by confocal spatial filter and to support it within framework of simple theoretical model.

**Basic Equations.**

In the scalar diffraction approximation the structure of the optical image is given by the following convolution equation/1,2/:

$$E_{image}(\vec{r}) = \iint E_{illum}(\vec{r}\,')\left[1 + N(\vec{r}\,')\right] f(\vec{r}\,') P(\vec{r}+\vec{r}\,') d^2\vec{r}\,'$$

where $P$ is so-called response function (optical transfer function) of the microscope, say propagator of the optical system (the argument of $P$, in form of $\vec{r}+\vec{r}\,'$, shows that the image after ocular is inverted compared to object), $f$ – transmittance of the object, the amplitude of $f$ is responsible for absorption, the phase of $f$ - for phase changes, $N$ - 2D random noise field, $E_{illum}$ - illuminating field after the condenser. This equation is obtained by considering the successive propagation of light through the sequence of optical elements, described as amplitude – phase screens. For example, in paraxial approximation the lens has transfer function in the form of imaginary exponent with parabolic phase profile $\exp(ik|\vec{r}|^2/2F)$, where $k$ is wavenumber $2\pi/\lambda$, $F$ – is the focal length, $\vec{r}$ - the vector in the plane perpendicular to direction of propagation. The diafragm has transfer function in the form of the regular complex function $D(\vec{r})$ of real variable $\vec{r}$, describing the variable transmission and phase corrections in the diafragm plane. The random phase and amplitude distortions of the light structure are taken in account by complex random field $N(\vec{r})$ of real variable $\vec{r}$ /2/.

In operator form we have :

$$E_{image}(\vec{r}) = \hat{K}\left\{\hat{P} E_{illum}(\vec{r}) \left[1 + N(\vec{r})\right] f(\vec{r}) \right\}, \quad (1)$$

where $\hat{P}$ is convolution integral describing OTA (optical transfer function) of the microscope, $\hat{K}$ - the similar integral, introduced for confocal spatial filter /2/(see **fig.1**) .

The noise random field $N(\vec{r})$ is considered here not as perturbation: its average value and moments could comparable or even larger than those of illuminating random field $E_{illum}(\vec{r})$. Both the illuminating source $E_{illum}$ and noise contamination field $N(\vec{r})$ are considered, usually, as $\delta$-correlated, statistically independent from each other random fields:

$$\langle E_{illum}(\vec{r}), E^{*}_{illum}(\vec{r}^{'}) \rangle = I_0(\vec{r})\delta(\vec{r}-\vec{r}^{'}) \quad ; \quad \langle N(\vec{r}), N(\vec{r}^{'}) \rangle = \delta(\vec{r}-\vec{r}^{'})$$

where $I_0(\vec{r})$ is the spatial distribution of intensity of light in the object plane /1/.

**Van Cittert-Zernicke theorem.**

The case of objective with circular aperture and ideal relay system had been intensively considered previously/2/. In this case the kernel of equation (1) or response function *P* has the following form:

$$P(\vec{r}+\vec{r}^{'}) = \frac{ik}{2\pi z} A \exp(ikz) \frac{J_1[k(\vec{r}+\vec{r}^{'})a/z]}{[k(\vec{r}+\vec{r}^{'})a/z]}$$

$k = 2\pi/\lambda$ is wavenumber, *a* – diameter of aperture, *z* –distance between object and image.

Now following to /1,2/ consider the second-order correlation function :

$$\Gamma(\vec{r},\vec{r}^{'}) = \langle E_{image}(\vec{r}) E^{*}_{image}(\vec{r}^{'}) \rangle$$

for optical field in the image plane:

$$\Gamma(r,\vec{r}^{'}) = \iint \langle E(r_1) E^{*}(\vec{r}_1^{'}) \rangle \left[1 + \langle N(\vec{r}_1) N^{*}(\vec{r}_1^{'}) \rangle \right] f(\vec{r}^{'}_1) f^{*}(\vec{r}_1) P(\vec{r}^{'}_1+\vec{r}^{'}) P^{*}(\vec{r}_1+\vec{r}) d^2\vec{r}_1 d^2\vec{r}_1^{'}$$

So we have correlation $\Gamma(\vec{r},\vec{r}^{'})$ function in the form of the sum of two components, regular:

$$\Gamma_{REG}(r,\vec{r}^{'}) = \iint \langle E(r_1) E^{*}(\vec{r}_1^{'}) \rangle f(\vec{r}^{'}_1) f^{*}(\vec{r}_1) P(\vec{r}^{'}_1+\vec{r}^{'}) P^{*}(\vec{r}_1+\vec{r}) d^2\vec{r}_1 d^2\vec{r}_1^{'}$$

and noisy one:

$$\Gamma_{NOISE}(r,\vec{r}^{'}) = \iint \langle E(r_1) E^{*}(\vec{r}_1^{'}) \rangle \left[ \langle N(\vec{r}_1) N^{*}(\vec{r}_1^{'}) \rangle \right] f(\vec{r}^{'}_1) f^{*}(\vec{r}_1) P(\vec{r}^{'}_1+\vec{r}^{'}) P^{*}(\vec{r}_1+\vec{r}) d^2\vec{r}_1 d^2\vec{r}_1^{'}$$

Rewriting these equations in operator form we have:

$$\Gamma(r,\vec{r}^{'}) = \hat{K}\hat{K}^{\bullet}\left\{ \hat{P}\hat{P}^{\bullet}\langle E_{illum}(\vec{r}_1) E^{*}_{illum}(\vec{r}_1^{'}) \rangle \left[1 + \langle N(\vec{r}_1) N(\vec{r}_1^{'}) \rangle\right] f(\vec{r}^{'}_1) f^{*}(\vec{r}_1) \right\} \quad (3),$$

where convolution operators $\hat{K}$, $\hat{K}^{\bullet}$ and $\hat{P}$, $\hat{P}^{\bullet}$ affect on different spatial variables , $\vec{r}^{'}, \vec{r}$ correspondingly. The $\delta$- correlation of illumination and distortion field random fields makes possible the substantial simplification of the equation (3). Firstly, in the absence of external noise source, i.e. when $N(\vec{r}) \equiv 0$, and without spatial filter, i.e. when $\hat{K} \equiv 1$ we have the Zernike theorem in its classical form :

$$\Gamma(r,\vec{r}\,') = \hat{P}\,\hat{P}^{\bullet}\left\{\, I_0(\vec{r}_1)\, f^*(\vec{r}_1)\, f(\vec{r}_1^{\,'})\,\right\}\,,$$

i.e. in coordinate representation :

$$\Gamma(r,\vec{r}\,') = \int I_0(\vec{r}_1)\, f^*(\vec{r}_1)\, f(\vec{r}_1^{\,'})\, P(\vec{r}_1+\vec{r})\, P^*(\vec{r}_1+\vec{r}\,')\, d^2\vec{r}_1\,,$$

for example, the intensity distribution immediately follows /2/:

$$I(r) = \int I_0(\vec{r}_1)\, |f^*(\vec{r}_1)|^2\, |P(\vec{r}_1+\vec{r})|^2\, d^2\vec{r}_1$$

**Confocal spatial filtering .**

The noise random field $N(\vec{r})$ affects the quality of image, introducing the significant spread of Fourier spatial spectrum. It is shown on the fig.2 by numerical modeling on 128 x 128 mesh as image of the object in the form of rectangular carpet is distorted by additive Rayleigh noise/2/. In our model we used quite general, additive model of noise/3/, taking into account both phase and amplitude distortions. The latter, for example, could be initiated by dust particles on surfaces of the microscope components, introducing $\delta$-like randomly located obstacles. The total intensity associated with noise term exceeded 100 times the intensity of the "carpet". The usage of additional confocal spatial filter enhanced filtering action of microscope itself . The image had been restored with correlation 90%, compared to initial image at the expense of signal losses at 50 % level.

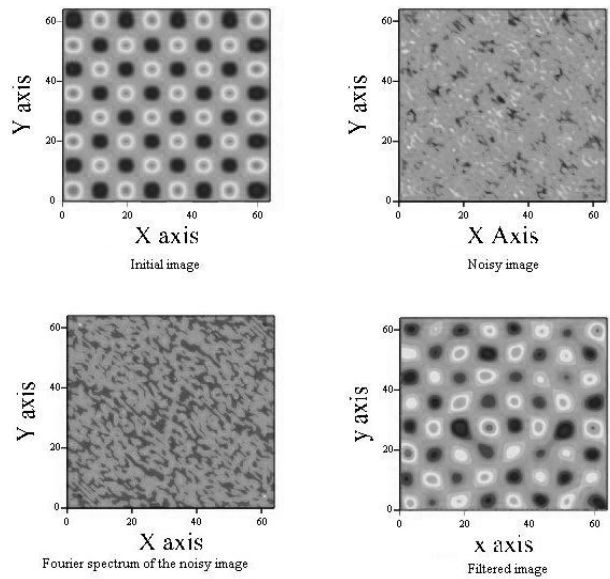

**Fig.2**

The qualitative agreement with results of Sheppard and Gauderon /4/ had been obtained. The signal to noise ratio *S/N* had been calculated using equation (3) on the 128 x 128 numerical mesh. The signal-to-noise ratio as a function of diaphragm width has maximum near boundary of the spatial spectrum of the signal. For the smaller diameters the *S/N* drops to zero linearly.

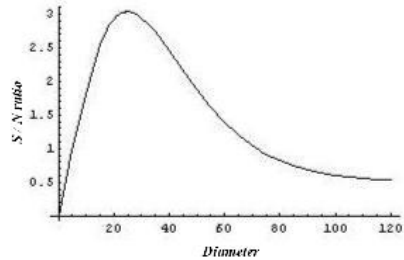

**Fig.3**

**Bioimaging with confocal spatial filtering and commercial digital camera.**

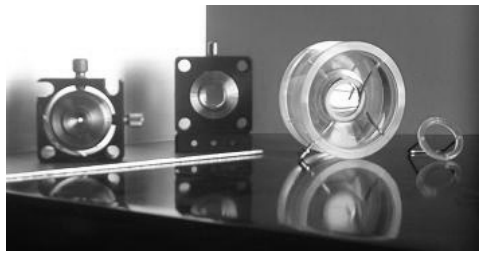

**Fig.4**

The inverted microscope with 10x objective had been improved by confocal filter for observation of hyppocampal slices. Fig.3 shows the general view of the microscope field (diameter of the working area 18 mm) through 5x lupa, obtained by HP Photosmart – 320 digital camera ( 2 Megapixel, maximum resolution 1200 x 1600 ) .

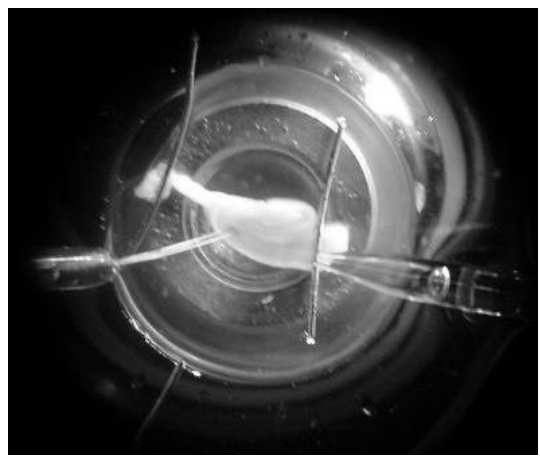

**Fig.5**

The further observations had been made with the 10x objective and *Genius WEB – cam* with resolution 300x200 pixels and manual focusing. The total field of view in present frames is approximately 400 μm (see **fig.6** below).

The two slices illuminated with different *UV* filters at *380* and *360 nm* are shown in **fig.6.**

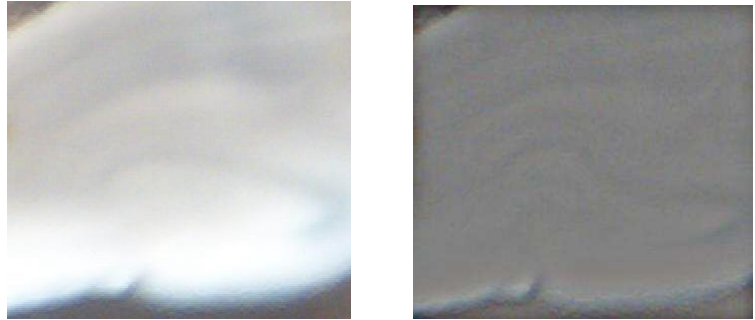

**Fig.6**

Thus we see that confocal spatial filtering provided reasonable image quality with commercial *CCD* – camera under incoherent continuous illumination.

**Acknoledgements.**

The author acknowledges Dra. M.E.O.Quinta-Ferreira for formulation of problem.
This work was partially supported by NATO ICCTI grant.